\documentclass{article}
\usepackage{spconf,amsmath,graphicx}
\usepackage{bm}
\usepackage{booktabs}
\usepackage{multirow}
\usepackage{amssymb}
\usepackage{bbding}
\usepackage{color}

\title{Learning Contextually Fused Audio-visual Representations for Audio-visual Speech Recognition}
\name{Zi-Qiang Zhang$^1$, Jie Zhang$^{1,3}$, Jian-Shu Zhang$^{2}$, Ming-Hui Wu$^1$, Xin Fang$^1$, Li-Rong Dai$^1$
}
\address{ $^1$NEL-SLIP, University of Science and Technology of China (USTC), Hefei, China \\
	$^2$iFlytek Research, iFlytek Co., Ltd., Hefei, China \\
	$^3$State Key Laboratory of Acoustics, Institute of Acoustics, Chinese Academy of Sciences, Beijing, China}

\begin{document}
\ninept
\maketitle
\begin{abstract}
	With the advance in self-supervised learning for audio and visual modalities, it has become possible to learn a robust audio-visual speech representation. 
	This would be beneficial for improving the audio-visual speech recognition (AVSR) performance, as the multi-modal inputs contain more fruitful information in principle.
	In this paper, based on existing self-supervised representation learning methods for audio modality, we therefore propose an audio-visual representation learning approach.
%
	The proposed approach explores both the complementarity of audio-visual modalities and long-term context dependency using a transformer-based fusion module and a flexible masking strategy.
	After pre-training, the model is able to extract fused representations required by AVSR.
	Without loss of generality, it can be applied to single-modal tasks, \textit{e.g.}, audio/visual speech recognition by simply masking out one modality in the fusion module.
%
	The proposed pre-trained model is evaluated on speech recognition and lipreading tasks using one or two modalities, where the superiority is revealed.
\end{abstract}
\section{Introduction}
\label{sec:intro}
Human perceives speech in a multimodal way, as looking at a talking face helps understand the corresponding voice~\cite{mcgurk1976hearing}.
The complementarity of visual and audio modalities has been widely studied and was shown to be useful for audio-visual speech recognition (AVSR) especially under noisy conditions~\cite{8461326Petridis,8585066Afouras,1145Sterpu,9306348Wei,9414567Ma}.
However, for AVSR a large amount of manually transcribed labels are required, and the calibration of these labels is rather time-consuming and expensive.
As unsupervised pre-training has been commonly-used in low-resource applications~\cite{HAN2021}, an intriguing question arises {\it whether we can learn audio-visual representations without any label, which can then be utilized for AVSR?}

There are a few works on unsupervised audio-visual representation learning. For example, the audio and visual representations can be jointly learned by audio-visual synchronization~\cite{Owens_2018_ECCV,8682524Chung,Korbar2018}, correspondence~\cite{Arandjelovic_2017_ICCV,NEURIPS2020Alwassel,ma2021contrastive,liu2021cross} and instance discrimination~\cite{Morgado_2021_CVPR1,Morgado_2021_CVPR2}.
One can also use one modality as targets to learn another modality~\cite{Owens2016Ambient,ma2021lira}.
Nevertheless, there are at least two challenges limiting these approaches applying to practical AVSR systems.
Firstly, most of them learn instance-level representations for short-term tasks such as sound event classification.
Such learning objective may not be suitable for speech recognition, which requires sequence-level representations that vary continuously at frame-level and contain a long-term context dependency.
Secondly, the audio and visual representations are extracted separately. Although the individual modal representations can be jointly optimized by the loss function, the lack of intermediate interactions restricts the information flow across modalities, as it was shown in~\cite{8585066Afouras,1145Sterpu,9306348Wei} that an early fusion mechanism benefits the AVSR performance.

Regarding the first challenge, a new training objective was proposed in~\cite{shukla2020learning}, where the audio embeddings are used to reconstruct the subsequent visual images over the time dimension. 
Audio-visual inpainting was designed in~\cite{9413488Morrone}, where the masked speech spectrogram is predicted using the visual information and audio contexts.
A versatile network presented in~\cite{ma2021contrastive} can be used to learn both global and local representations.
Moreover, the masked prediction~\cite{devlin2018bert} was shown to be  surprisingly effective to learn context dependency in~\cite{NEURIPS2020Baevski,hsu2021hubert}, though the focus is merely on the audio modality.
For the second one, several fusion strategies have been proposed, including cross-modal attention~\cite{Cheng2020} and unimodal BERT~\cite{lee2020parameter}.

Inspired by~\cite{NEURIPS2020Baevski,9413488Morrone,ma2021contrastive}, in this paper we propose a new self-supervised audio-visual representation learning approach for AVSR.
It can be seen as a multi-modal extension of Wav2vec2.0~\cite{NEURIPS2020Baevski}.
We extend the audio-visual correspondence and masked prediction into a unified learning objective: audio-visual recovery, where the model is trained to recover the corrupted information of one modality using 1) the complementarity from the other modality and 2) the contexts in a sequence.
In detail, we learn three types of embeddings, namely audio embeddings, visual embeddings and fused embeddings.
The fused embeddings are extracted by a fusion module exploring cross-modal and context information and then used to recover the corrupted audio \& visual embeddings at the frame level.
The recovery is evaluated using a contrastive loss~\cite{NEURIPS2020Baevski}, where the target embeddings are treated as positive samples.
The corruption can be synthesized via data augmentation like temporal masking.
Compared to the single-modal mask, the considered masking strategy for multi-modalities is more flexible to control the priority of the information that accessed by the model (detailed in Section~\ref{sec:method}).
For the downstream task, the fused embeddings are directly used for AVSR.
We find that the pre-trained model is also able to cope with single-modal tasks like automatic speech recognition (ASR), visual speech recognition (VSR) and lipreading~\cite{9053841Martinez}, by simply masking out the irrelevant modality before fusion.

\begin{figure*}[!t]
	\begin{minipage}[b]{0.36\linewidth}
		\centering
		\centerline{\includegraphics[width=\columnwidth]{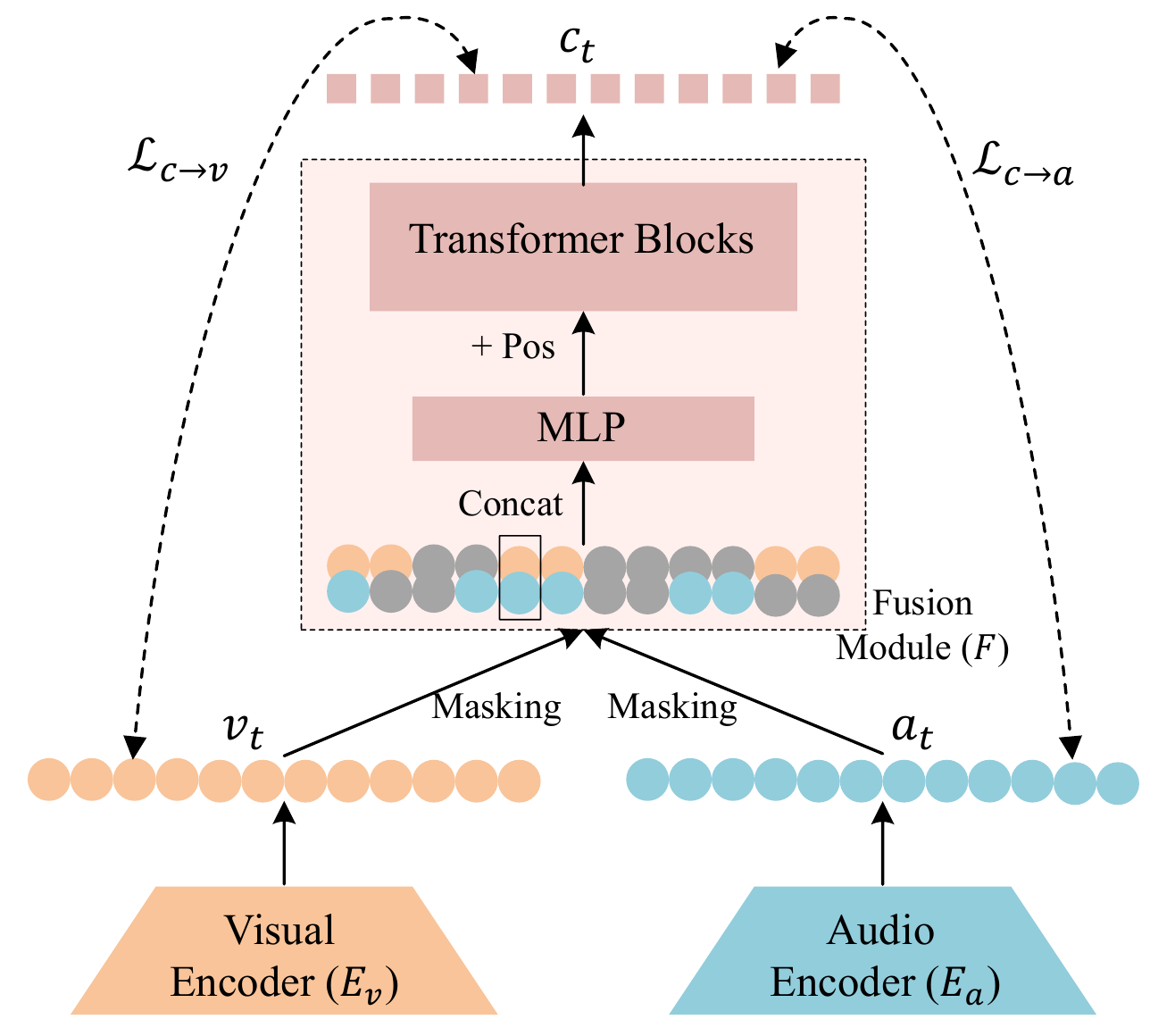}}
		\centerline{(a) Audio-visual Pre-training}\medskip
	\end{minipage}
	\begin{minipage}[b]{0.198\linewidth}
		\centering
		\centerline{\includegraphics[width=\columnwidth]{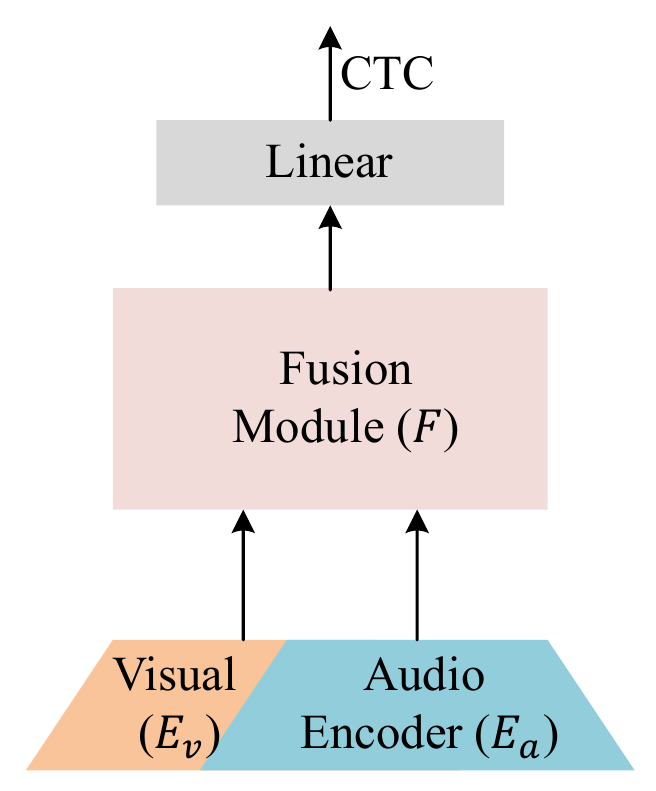}}
		\centerline{(b) AVSR}\medskip
	\end{minipage}
	\begin{minipage}[b]{0.18\linewidth}
		\centering
		\centerline{\includegraphics[width=\columnwidth]{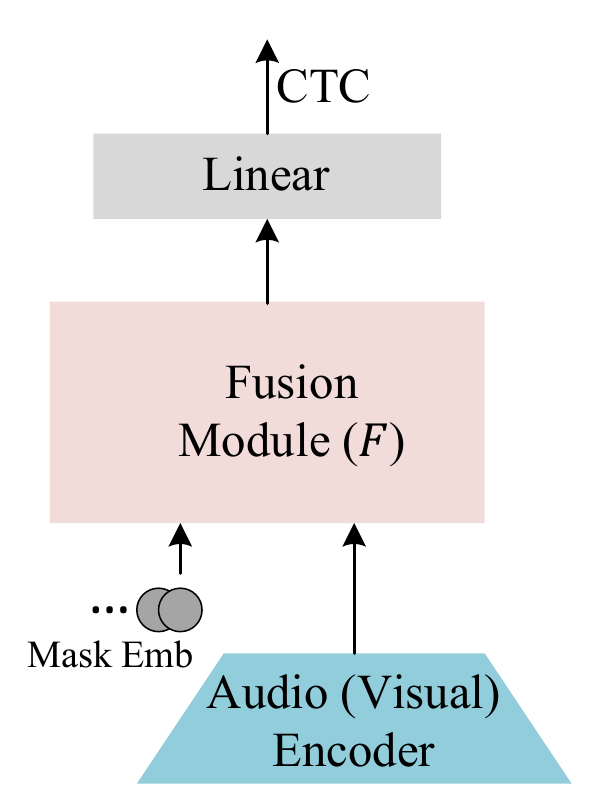}}
		\centerline{(c) ASR or VSR}\medskip
	\end{minipage}
	\begin{minipage}[b]{0.18\linewidth}
		\centering
		\centerline{\includegraphics[width=\columnwidth]{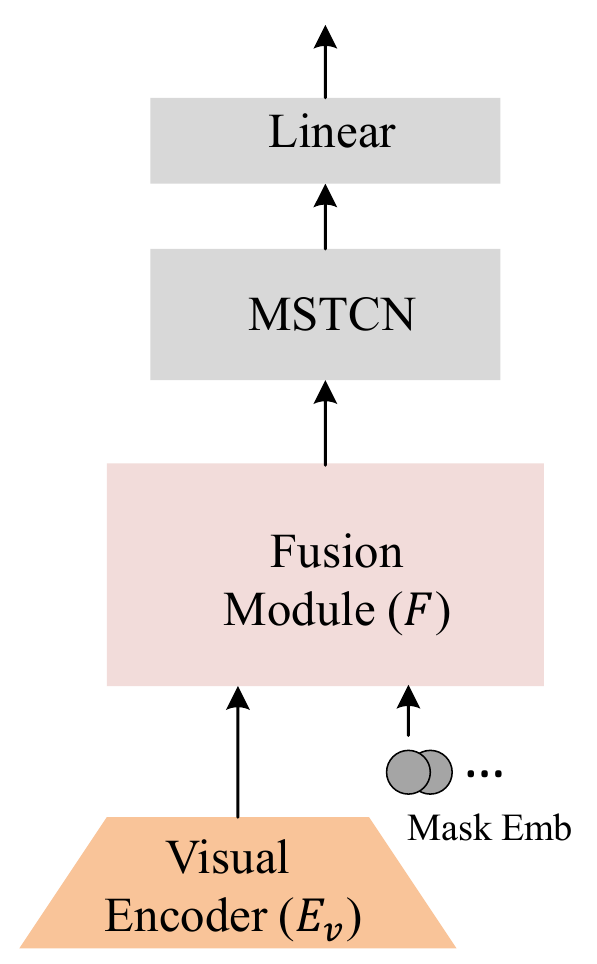}}
		\centerline{(d) Lipreading}\medskip
	\end{minipage}
	\vspace{-0.2cm}
	\caption{An illustration of the proposed method: (a) the proposed audio-visual representation learning framework, (b) application to AVSR using pre-trained modules, (c) application to ASR (or VSR) where the visual encoder (or the audio encoder) is excluded, and (d) application to Lipreading, where the audio encoder is excluded and a MSTCN module is added.}
	\label{fig:1}
\end{figure*}

Recently, we found that \cite{shi2022learning} also leveraged masked prediction to learn audio-visual representations, which was concurrent with this work.
They use the HuBERT \cite{hsu2021hubert} clusters as the pre-training targets, instead, we employ contrastive learning in this paper.
Their model size and training configurations are quite different with ours so the results are not compared in this paper.

Our contributions are summarized as three folds, 1) we propose a self-supervised framework to learn audio-visual representations for AVSR which can fuse the information of two modalities as well as their temporal context 2) We show the fused representation is also capable and flexible to copy with single-modal applications; 3) Experiments demonstrate that our pre-trained model significantly improves the performance of AVSR, ASR and VSR, and achieve the state-of-the-art performance on LRW~\cite{Chung2017LRW}.

\section{Methods} \label{sec:method}
\subsection{Problem formulation}	\label{ssec:2.1}
Suppose we have a video dataset containing synchronized audio and visual recordings, for example $\bm{x}^a=\{\bm{x}_1^a,\bm{x}_2^a,\cdots,\bm{x}_N^a\}$ being a sequence of audio sampling points and $\bm{x}^v=\{\bm{x}_1^v,\bm{x}_2^v,\cdots,\bm{x}_M^v\}$ being a series of images.
The length of $\bm{x}^a$ and $\bm{x}^v$ depends on their sampling rates.
Typically, self-supervised methods aim to learn audio and visual embeddings as
\begin{equation}
	\bm{a}=\{\bm{a}_1,\bm{a}_2,\cdots,\bm{a}_T\}=E_a(\bm{x}^a)	\label{eqn:Ea},
\end{equation}
\begin{equation}
	\bm{v}=\{\bm{v}_1,\bm{v}_2,\cdots,\bm{v}_T\}=E_v(\bm{x}^v)	\label{eqn:Ev},
\end{equation}
where $E_a$ ($E_v$) denotes the audio (visual) encoder and $T$ is the downsampled length.
The learning objective is built upon these two embeddings, \textit{i.e}.,
\begin{equation}
\mathcal{L} = sim(\bm{x}^a, \bm{x}^v),
\end{equation}
where the loss could be implemented by audio-visual synchronization~\cite{Owens_2018_ECCV}, correspondence~\cite{Arandjelovic_2017_ICCV} or contrastive loss~\cite{ma2021contrastive}, etc.

The goal of this work is to learn an extra embedding, say $\bm{c}$, which is called the contextually fused representation.
$\bm{c}$ fuses the audio and visual representation as well as their temporal contexts as
\begin{equation}
\bm{c}_t=F\left( \bm{a}_t,\bm{v}_t | \bm{a}, \bm{v} \right).
\end{equation}
Next, we will design a network (containing $E_a$, $E_v$ and $F$) and a learning objective to learn the contextually fused representation $\bm{c}$.

\subsection{Proposed audio-visual representation learning}	\label{ssec:2.2}
The proposed audio-visual representation learning framework is illustrated in Fig.~\ref{fig:1}(a).
The audio and visual embeddings are first extracted from $E_a$ and $E_v$ and dowmsampled to the same length $T$.
Then, masking is applied independently to the audio and visual embeddings before inputting to the fusion module as
\begin{equation}
\bm{\bar{a}}_{t}=\left\{\begin{aligned}
			\bm{a}_{t} &, t \notin M_{a} \\
			\bm{m} &, t \in M_{a}
\end{aligned}\right.,
\bm{\bar{v}}_{t}=\left\{\begin{aligned}
			\bm{v}_{t} &, t \notin M_{v} \\
			\bm{m} &, t \in M_{v}
\end{aligned}\right.
\end{equation}
\begin{equation}
\bm{c}_t=F\left( \bm{\bar{a}}_{t},\bm{\bar{v}}_{t} | \bm{\bar{a}}, \bm{\bar{v}} \right),
\end{equation}
where $M_a$ and $M_v$ are the mask positions of two modalities and $\bm{m}$ is the learnable mask embedding.
Note that the learning behavior is flexible as the masking can be randomly performed on two modalities.
For example, at timestep $t$, if $\bm{a}_t$ is masked while $\bm{v}_t$ is not, then the model can access $\bm{v}_t$ to recover $\bm{a}_t$;
in case both $\bm{a}_t$ and $\bm{v}_t$ are masked, $\bm{c}_t$ will be more dependent on their contexts, \textit{i.e.} the unmasked part.

For the fusion module $F$, we use one-layer multi-layer perceptron (MLP) followed by transformer blocks.
Due to the fact that the audio and visual embeddings are already synchronized at the frame level, arranging them along time as in~\cite{lee2020parameter} will lose the alignment information in subsequent transformer blocks, we therefore concatenate them in the feature dimension.
Finally, the learning objective is built for the masked frames of $\bm{a}_t$, $\bm{v}_t$ and $\bm{c}_t$ as

\begin{equation}
\mathcal{L} = \mathcal{L}_{c\rightarrow a} + \mathcal{L}_{c\rightarrow v},
\label{eqn:L}
\end{equation}
where
\begin{equation}
\mathcal{L}_{c \rightarrow a}=\sum\limits_{t \in M_{a}} \frac{\exp ^{\left\langle \bm{c}_{t}, \bm{a}_{t}\right\rangle}}{\exp ^{\left\langle \bm{c}_{t}, \bm{a}_{t}\right\rangle}\!+\!\sum\limits_{i \in M_{a}, i \neq t}\!\exp ^{\left\langle\bm{c}_{t}, \bm{a}_{i}\right\rangle}\!+\!\sum\limits_{i \neq t}\!\exp ^{\left\langle \bm{c}_{t}, \bm{v}_{i}\right\rangle}},
\label{eqn:La}
\end{equation}
\begin{equation}
\mathcal{L}_{c \rightarrow v}=\sum\limits_{t \in M_{v}} \frac{\exp ^{\left\langle \bm{c}_{t}, \bm{v}_{t}\right\rangle}}{\exp ^{\left\langle \bm{c}_{t}, \bm{v}_{t}\right\rangle}\!+\!\sum\limits_{i \in M_{v}, i \neq t}\!\exp ^{\left\langle \bm{c}_{t}, \bm{v}_{i}\right\rangle}\!+\!\sum\limits_{i \neq t}\!\exp ^{\left\langle \bm{c}_{t}, \bm{a}_{i}\right\rangle}},
\label{eqn:Lv}
\end{equation}
which are contrastive losses of two modalities for one input sequence.
Specifically, $\mathcal{L}_{c\rightarrow a}$ and $\mathcal{L}_{c\rightarrow v}$ treat $\bm{a}_t$ and $\bm{v}_t$ as positive samples repectively and other masked embeddings as negative samples.
$\left\langle \cdot \right\rangle$ measures the cosine similarity of two embeddings.
We also find that the inclusion of cross-modal negative samples (\textit{i.e.}, the last term of the denominator in~(\ref{eqn:La}) or~(\ref{eqn:Lv}), see Section~\ref{ssec:4.1}) is important, as it can push $\bm{a}_t$, $\bm{v}_t$ and $\bm{c}_t$ into the same embedding space, leading to a unified representation.

\subsection{Applications of the pre-trained model}	\label{ssec:2.3}
The pre-trained model can be easily applied to downstream tasks. For example, the application to AVSR as shown in Fig.~\ref{fig:1}(b) is rather straightforward.
Without loss of generality, the learned modules can also handle single-modal tasks, where only the audio or visual data are available.
One can simply replace the visual embedding with the learned mask embedding ($\bm{m}$) for the ASR task  as shown in Fig.~\ref{fig:1}(c).
On the contrary, the audio encoder could be replaced by the visual encoder for VSR and lipreading as in Fig.~\ref{fig:1}(c) and Fig.~\ref{fig:1}(d).
Following the fine-tuning strategy in~\cite{NEURIPS2020Baevski,ma2021lira}, the pre-trained modules are jointly tuned with additional task-specific layers.

\section{Experimental Setup} \label{sec:exp}
\subsection{Pre-training}	\label{ssec:3.1}
We use the \textit{pretrain} subset of LRS3~\cite{afouras2018lrs3} for pre-training.
It contains about 408 hours of video speech with an average utterance duration of 12.4 seconds.
We refer the reader to~\cite{afouras2018lrs3} for more details.
The audio inputs are raw waveforms extracted from the original video.
For visual modality, we first crop $112\times112$ pixels from the center and then average the RGB channels to gray images.

The audio encoder $E_a$ is a stack of 1D-CNN layers, which can be found in~\cite{Schneider2019}.
We add one more convolutional layer with a kernel size of 2 and stride of 2 on the top of the 1D-CNN layers, so that the audio embeddings can match the video sampling rate in LRS3 (40 ms per frame).
The visual encoder $E_v$ consists of a 3-D CNN layer followed by a Resnet-18~\cite{He_2016_CVPR} block, and the details can be found in~\cite{9053841Martinez}.
The fusion module consists of a {MLP with single hidden layer} and 12 transformer~\cite{Vaswani2017} blocks, where the MLP has 2048 hidden units and each transformer block has an attention dimension of 512 and a feed-forward dimension of 2048.
The dimensions of all three embeddings are 512, which are then linearly reduced to 128 for computing the loss.
As such, the size of all parameters of the proposed pre-training model is around 60 M.

The masking strategy keeps the same as~\cite{NEURIPS2020Baevski} with a masking probability of 0.65 and a window size of 5 (200 ms) for both audio and visual embeddings.
Considering that the visual encoder has a temporal receptive field of 5, we automatically expand the true visual mask boundaries to avoid information leakage using adjacent frames.
In all pre-training experiments, we sample a fixed number of negative samples from the masked embeddings, which is set to be 100.
In case the cross-modal sampling is enabled, 50 negatives are sampled from both audio and visual embeddings, respectively.

Furthermore, the Adam optimizer is used in training with a constant learning rate of 5e-4.
Training is conducted on 32 GPUs and the batch size on each GPU is up to 100 seconds. We simulate such number of GPUs by gradient accumulation using Fairseq~\cite{ott2019fairseq}.
We discard all samples that are shorter than 1 s (\textit{i.e.}, 25 frames) and crop to be a maximum length of 20 s (\textit{i.e.}, 500 frames) for longer samples.
The full number of training iterations is 300 K, and for ablation studies it is set to be 100 K as longer training does not change the relative performance gap between different models.

\subsection{Downstream tasks}	\label{ssec:3.2}
\subsubsection{AVSR}	\label{ssec:3.2.1}
As a smaller dataset can better reflect the effectiveness of a pre-trained model, we use the \textit{trainval} subset of LRS3 for AVSR.
It contains about 30 hours of labeled training data, and we randomly split out $5\%$ of them as the \textit{validation} set.
The data processing keeps the same as in Section~\ref{ssec:3.1}.
For tuning, a linear classifier is added on the top of the fusion module (see Fig.~\ref{fig:1}(b)) and the character-level CTC~\cite{graves2006connectionist} loss is computed.
The temporal masking is employed here as data augmentation.
The model is trained for 25 K steps with a batch size of 100 seconds per GPU on 16 GPUs.
We use the scheduled learning rate with a warning up, a holding and an exponentially decaying period.
The maximum learning rate is $5\times10^{-4}$ and the period ratio is $(0.2,0.3,0.5)$.
After tuning, we chose the checkpoint with the best performance on the \textit{validation} set and evaluate the character error rate (CER) as well as the word error rate (WER) by \textit{Viterbi} decoding without language model on the 0.4 version of the \textit{test} set.

\subsubsection{ASR \& VSR}	\label{ssec:3.2.2}
For the application to ASR or VSR, the data, model, loss and training schedule are all the same as AVSR. The only difference lies in that we only keep a single modality and exclude the other (see Section~\ref{ssec:2.3}).
In order to avoid overfitting, the number of training steps for VSR is shortened to be 10 K.

\subsubsection{Lipreading}	\label{ssec:3.2.3}
We use the LRW~\cite{Chung2017LRW} dataset for the lipreading task, where the objective is to classify spoken words according to a short segment of facial images.
LRW contains about 173 hours for 500 spoken words (157 hours for training), where each record is an 1.16-second clip.
The considered fine-tuning model is illustrated in Fig.~\ref{fig:1}(d).
As it was shown that a backend network aggregating frame-level information is necessary for word-level classifications~\cite{9053841Martinez}, we add a MSTCN~\cite{9053841Martinez} module on the top of the fusion module.
Then, the outputs of MSTCN are averaged over time, followed by a linear classification layer.
Following~\cite{9053841Martinez}, we preprocess the images by extracting $96\times96$ pixels of mouth regions of interest (ROIs), and then randomly crop to $88\times88$ for training.
For data augmentation, we use random flipping, mixup~\cite{zhang2017mixup} with the Beta weighting distribution $\omega \sim Beta (0.4,0.4)$, and temporal masking with a fixed mask length of $9$ on the output of the visual encoder.
The training is done on 16 simulated GPUs and the batch size on each GPU is 100.
The learning rate schedule is the same as in the AVSR task.

\begin{table*}[htb]
	\centering
	\caption{Performance on ASR, VSR and AVSR tasks. Models are fine-tuned on 30 hours of labeled data after pre-training.}
	\begin{tabular}{cccc|cc|cc|cc@{}}
		\toprule
		\multirow{2}{*}{Exp}	& \multicolumn{1}{c}{\multirow{2}{*}{Pre-training Loss}}		& Cross-modal	& Pre-training		& \multicolumn{2}{c|}{ASR}	& \multicolumn{2}{c|}{AVSR}	& \multicolumn{2}{c}{VSR}	\\
		&																& Negatives		& Modality	 		& CER(\%)			& WER(\%)		& CER(\%) 		& WER(\%)		& CER(\%)		& WER(\%)       \\ \midrule
		B0	& --																	& --			& --				& 7.2				& 19.2			& 7.1			& 18.1			& 65.4 			& 99.4			\\
		B1	& $\mathcal{L}_{c \rightarrow v}$ 										& --			& Visual            & --                & --            & --            & --            & 67.9          & 112.0         \\
		B2	& $\mathcal{L}_{c \rightarrow a}$ (Wav2vec2.0)                			& --			& Audio             & 5.1               & 14.0          & --            & --            & --            & --            \\ \midrule
		C1	& $\mathcal{L}_{c \rightarrow v}$                        				& $\times$		& Audio \& Visual   & 6.3               & 17.6          & 5.8           & 16.4          & 52.3          & 88.6          \\
		C2	& $\mathcal{L}_{c \rightarrow a}$                       				& $\times$		& Audio \& Visual   & 4.9               & 13.6          & 4.2           & 12.1          & 51.1          & 88.0          \\ \midrule
		D1	& $\mathcal{L}_{c\rightarrow a}+\mathcal{L}_{c\rightarrow v}$			& $\times$		& Audio \& Visual   & 4.8               & 13.3          & 4.2           & 12.5          & 42.5          & 75.3          \\
		D2	& $\mathcal{L}_{c\rightarrow a}+\mathcal{L}_{c\rightarrow v}$			& $\checkmark$	& Audio \& Visual   & 4.6               & 12.7          & 4.0           & 11.6          & 42.0          & 74.1          \\
		D3	& D2 @300k 	& $\checkmark$	& Audio \& Visual   & 3.8               	& 10.9          & 3.1           & 9.1           & 38.5          & 67.8          \\ \bottomrule
	\end{tabular}
	\label{Tab:general}
	\vspace{-0.3cm}
\end{table*}

\section{Results} \label{sec:Results}

\subsection{Continuous speech recognition} \label{ssec:4.1}
At first, we evaluate the pre-trained models on the speech recognition, and the results are shown in Table~\ref{Tab:general}.
In Table 1, B0 (\textit{i.e.}, the first line) indicates a randomly initialized baseline.
B1 and B2 are single-modal baselines, where we exclude one modality according to Section~\ref{ssec:2.3} throughout pre-training and fine-tuning.
In fact, the audio baseline is the same as Wav2vec2.0~\cite{NEURIPS2020Baevski} if we ignore the MLP in our fusion model and the quantizer in Wav2vec2.0.
The rest results are obtained by the proposed method.

From Table 1, comparing B1 and B2 we can see that the audio pre-trained model can significantly improve the ASR performance than the visual model on VSR. 
It is also observed by comparing C1 with C2 that the pre-training loss $\mathcal{L}_{c \rightarrow a}$ works better than $\mathcal{L}_{c \rightarrow v}$.
These imply that the audio signal is more informative than the visual signal for learning speech representations.
Then, pre-training both $E_a$ and $E_v$ (C1, C2) allows the model to cope with the AVSR task, which reduces the WER by up to 13.6\% relatively compared to the single-modal baseline (\textit{i.e.}, 12.1 vs 14.0).
Moreover, on single-modal tasks (C1~vs.~B1 on VSR, C2~vs.~B2 on ASR), the audio-visual pre-trained models can also outperform the single-modal baselines.
This means that fusing an auxiliary modality at the pre-training stage always helps representation learning, even such a modality is not used in downstream tasks.
It also implies that the learned mask embedding is capable to replace a whole modality when applying to a signal-modal task.
This effect is further demonstrated in Table~\ref{Tab:mask}, where we compare the performance of an audio-visual pre-trained model on ASR using different methods to exclude the visual modality.
It shows that replacing with the learned mask embedding can best alleviate the influence of discarding visual inputs.

\begin{table}[!t]
	\centering
	\renewcommand\arraystretch{1.15}
	\caption{Different visual modality exclusion strategies for ASR with the AVSR topline in the last column. $\bm{x}^v$, $\bm{v}$ and $\bm{m}$ indicate visual input, visual embedding and the learned mask embedding.}
	\begin{tabular}{@{}c|cccc@{}}
		\hline
		Fine-tuning strategy & $\bm{x}_t^v\rightarrow\bm{0}$ 	& $\bm{v}_t\rightarrow\bm{0}$ 	& $\bm{v}_t\rightarrow\bm{m}$ 	& -- 	\\ \hline
		WER(\%)              & 13.8								& 13.0 							& 12.7							& 11.6		\\ \hline
	\end{tabular}%
	\label{Tab:mask}
\end{table}

\begin{figure}[!t]
	\centering
	\includegraphics[width=\columnwidth]{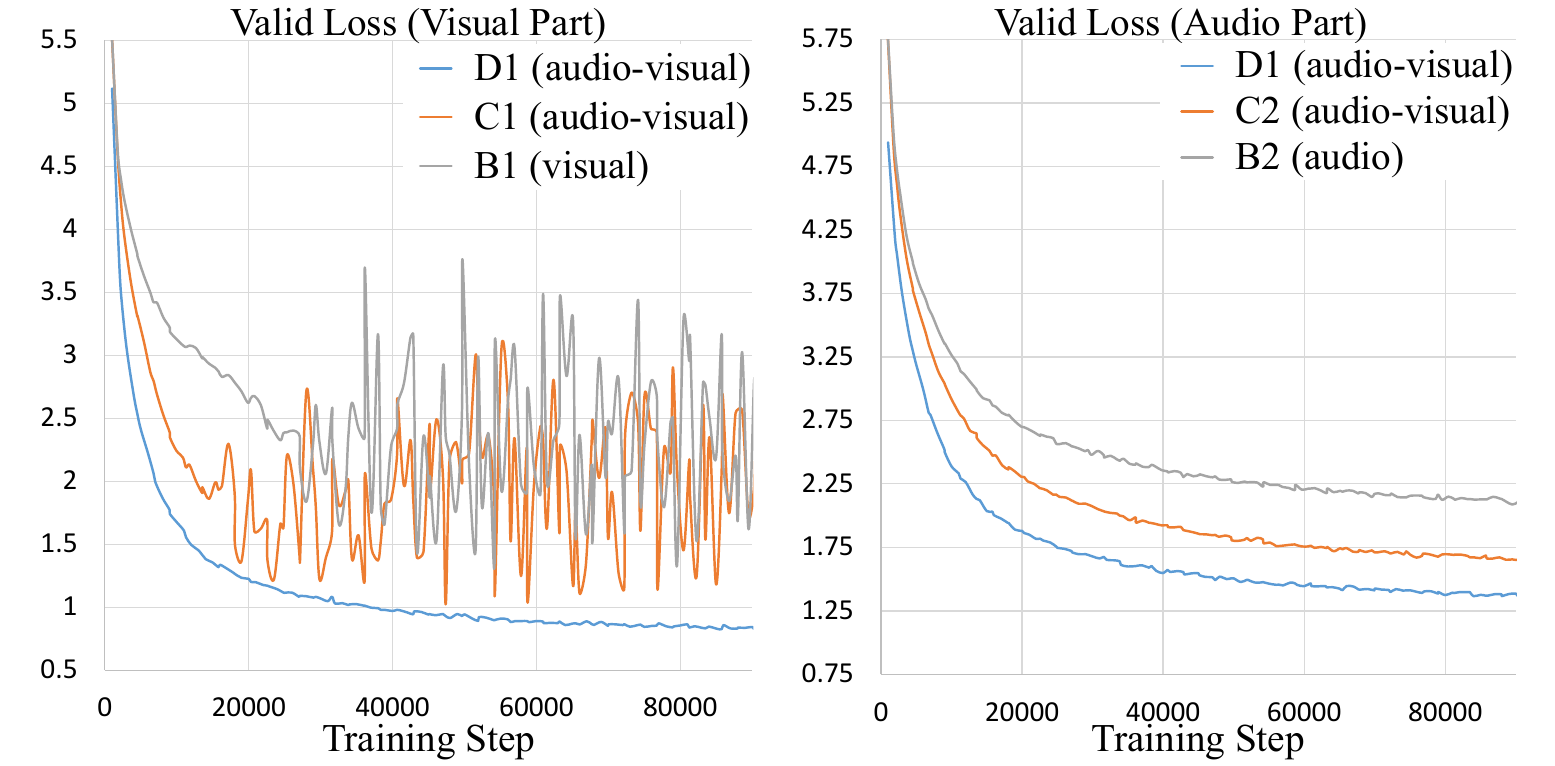}
	\caption{The loss curves on the \textit{validation} set using different training objectives and modalities with the same masking strategy and the same number of negatives.}
	\label{fig:loss}
	\vspace{-0.3cm}
\end{figure}

Moreover, from Table~\ref{Tab:general} it is clear that jointly optimizing $\mathcal{L}_{c\rightarrow a}$ and $\mathcal{L}_{c\rightarrow v}$ (D1) can further improve the VSR performance significantly (D1~vs.~C1/C2) and the ASR performance slightly.
This can be interpreted by observing the loss curves trained with different loss functions, which are shown in Fig.~\ref{fig:loss}.
In fact, using the combined loss turns out to be two interactions between two modalities, \textit{i.e.}, the explicit modality fusion at the feature level and the implicit connection at the loss level.
The former is performed by the fusion module, and this is why the loss of C1 is lower that B1 (also C2 lower than B2) in Fig.~\ref{fig:loss}.
The latter helps the knowledge distilled from one modality to the other, which somehow provides targets for each other.
Due to the distillation enabled by the combination of loss functions, D1 can stabilize $\mathcal{L}_{c\rightarrow v}$ when training.
This is more helpful for the visual modality due to the audio dominance.
Further, it can be seen that the addition of cross-modal negative samples (D2) improves the performance on all three tasks.
Such an operation would make the distillation easier, as directly comparing the embeddings of different modalities would push them tighter in a unified embedding space.
Besides, on the basis of D2 increasing the number of iterations for pre-training to be 300 K (i.e., D3) enables a further performance improvement.

%

\subsection{Lipreading}	\label{ssec:4.2}
Finally, we evaluate the proposed pre-trained model on the lipreading task in comparison with state-of-the-art (SOTA) methods.
The results are shown in Table~\ref{Tab:lipreading}.
Our supervised baseline uses the same model as our fine-tuning model, but is trained from scratch.
Compared to the vanilla MSTCN network~\cite{9053841Martinez}, the proposed supervised baseline includes an additional transformer module and the rest keeps the same from the view point of network structure.
It is clear that for lipreading the proposed self-supervised method increases the accuracy by 3.1\% compared to its supervised counterpart.
More importantly, it outperforms the SOTA self-supervised methods.

\begin{table}[!t]
	\centering
	\caption{The performance comparison on the lipreading task.}
		\resizebox{\columnwidth}{!}{%
		\renewcommand\arraystretch{1.0}
		\begin{tabular}{llc}
			\hline
			Methods			    						& Data (+ Pre-training Data)      						& Acc(\%) 		\\ \hline
			\multicolumn{3}{l}{\textit{Supervised}}                                                     		\\ \hline
			MSTCN~\cite{9053841Martinez}     			& LRW (157h)							 			& 85.3     		\\
			LiRA w/o pre-training~\cite{Ma9415063}     	& LRW (157h)							 			& 87.4     		\\
			DC-TCN~\cite{Ma_2021_WACV}					& LRW (157h)										& 88.4			\\
			Our baseline			                	& LRW (157h)										& 85.8     		\\ \hline
			\multicolumn{3}{l}{\textit{Self-supervised}}                                                     				\\ \hline
			Global \& Local~\cite{ma2021contrastive}	& LRW (157h) + LRS3,2 (632h)	      				& 86.9    		\\
			LiRA~\cite{ma2021lira}						& LRW (157h) + LRS3 (438h)		      				& 88.1    		\\
			Ours										& LRW (157h) + LRS3 (408h)      					& 88.9     		\\ \hline
		\end{tabular}%
		}
	\label{Tab:lipreading}
	\vspace{-0.2cm}
\end{table}

\section{Conclusion}	\label{ssec:5}
In this paper, we proposed a self-supervised audio-visual representation learning approach, which can explore the complementarity of two modalities and the temporal context dependency.
Evaluations on ASR, VSR, AVSR and lipreading tasks show that the proposed method enables better fused representations and is still effective in the absence of one modality, making it flexible for various applications.
We find that the audio modality dominates in the audio-visual self-supervised learning, which might lead to an imbalance between two modalities.
This can potentially be solved by training with a larger amount of data or applying a stronger augmentation on audio/visual inputs, which will be left as a part of our future research.

\vfill\pagebreak

\bibliographystyle{IEEEtran}
\bibliography{refs}

\end{document}